\documentclass[11pt]{article}
\usepackage{amsfonts}
\usepackage{amssymb}
\usepackage{graphics,psboxit,amsmath}
\usepackage{subfigure}
\usepackage{graphicx}
\usepackage{verbatim}
\usepackage{epic}
\usepackage{slashed}

\def\hybrid{\topmargin 0pt \oddsidemargin 0pt 
        \headheight 0pt \headsep 0pt
        \textwidth 16,5cm 
        \textheight 23cm 
        \marginparwidth .875in
        \parskip 5pt plus 1pt \jot = 1.5ex}

\hybrid

\catcode`\@=11

\def\marginnote#1{}
\newcount\hour
\newcount\minute
\newtoks\amorpm
\hour=\time\divide\hour by60
\minute=\time{\multiply\hour by60 \global\advance\minute by-\hour}
\edef\standardtime{{\ifnum\hour<12 \global\amorpm={am}%
        \else\global\amorpm={pm}\advance\hour by-12 \fi
        \ifnum\hour=0 \hour=12 \fi
        \number\hour:\ifnum\minute<10 0\fi\number\minute\the\amorpm}}
\edef\militarytime{\number\hour:\ifnum\minute<10 0\fi\number\minute}

\def\draftlabel#1{{\@bsphack\if@filesw {\let\thepage\relax
   \xdef\@gtempa{\write\@auxout{\string
      \newlabel{#1}{{\@currentlabel}{\thepage}}}}}\@gtempa
   \if@nobreak \ifvmode\nobreak\fi\fi\fi\@esphack}
        \gdef\@eqnlabel{#1}}
\def\@eqnlabel{}
\def\@vacuum{}
\def\draftmarginnote#1{\marginpar{\raggedright\scriptsize\tt#1}}

\def\draft{\oddsidemargin -.5truein
        \def\@oddfoot{\sl preliminary draft \hfil
        \rm\thepage\hfil\sl\today\quad\militarytime}
        \let\@evenfoot\@oddfoot \overfullrule 3pt
        \let\label=\draftlabel
        \let\marginnote=\draftmarginnote
   \def\@eqnnum{(\theequation)\rlap{\kern\marginparsep\tt\@eqnlabel}%
\global\let\@eqnlabel\@vacuum} }

\def\draft2{
        \def\@oddfoot{\sl preliminary draft \hfil
        \rm\thepage\hfil\sl\today\quad\militarytime}
        \let\@evenfoot\@oddfoot \overfullrule 3pt
        \let\label=\draftlabel
        \let\marginnote=\draftmarginnote
   \def\@eqnnum{(\theequation)\rlap{\kern\marginparsep\tt\@eqnlabel}%
\global\let\@eqnlabel\@vacuum} }

\def\preprint{\twocolumn\sloppy\flushbottom\parindent 2em
        \leftmargini 2em\leftmarginv .5em\leftmarginvi .5em
        \oddsidemargin -.5in \evensidemargin -.5in
        \columnsep .4in \footheight 0pt
        \textwidth 10.in \topmargin -.4in
        \headheight 12pt \topskip .4in
        \textheight 6.9in \footskip 0pt
        \def\@oddhead{\thepage\hfil\addtocounter{page}{1}\thepage}
        \let\@evenhead\@oddhead \def\@oddfoot{} \def\@evenfoot{} }

\def\numberbysection{\@addtoreset{equation}{section}
        \def\theequation{\thesection.\arabic{equation}}}

\def\underline#1{\relax\ifmmode\@@underline#1\else
        $\@@underline{\hbox{#1}}$\relax\fi}

\def\titlepage{\@restonecolfalse\if@twocolumn\@restonecoltrue\onecolumn
     \else \newpage \fi \thispagestyle{empty}\c@page\z@
        \def\thefootnote{\fnsymbol{footnote}} }

\def\endtitlepage{\if@restonecol\twocolumn \else \newpage \fi
        \def\thefootnote{\arabic{footnote}}
        \setcounter{footnote}{0}} 

\catcode`@=12
\relax

\def\figcap{\section*{Figure Captions\markboth
        {FIGURECAPTIONS}{FIGURECAPTIONS}}\list
        {Figure \arabic{enumi}:\hfill}{\settowidth\labelwidth{Figure
999:}
        \leftmargin\labelwidth
        \advance\leftmargin\labelsep\usecounter{enumi}}}
 \relax
\def\tablecap{\section*{Table Captions\markboth
        {TABLECAPTIONS}{TABLECAPTIONS}}\list
        {Table \arabic{enumi}:\hfill}{\settowidth\labelwidth{Table
999:}
        \leftmargin\labelwidth
        \advance\leftmargin\labelsep\usecounter{enumi}}}
 \relax
\def\reflist{\section*{References\markboth
        {REFLIST}{REFLIST}}\list
        {[\arabic{enumi}]\hfill}{\settowidth\labelwidth{[999]}
        \leftmargin\labelwidth
        \advance\leftmargin\labelsep\usecounter{enumi}}}
 \relax
\makeatletter
\newcounter{pubctr}
\def\publist{\@ifnextchar[{\@publist}{\@@publist}}
\def\@publist[#1]{\list
        {[\arabic{pubctr}]\hfill}{\settowidth\labelwidth{[999]}
        \leftmargin\labelwidth
        \advance\leftmargin\labelsep
        \@nmbrlisttrue\def\@listctr{pubctr}
        \setcounter{pubctr}{#1}\addtocounter{pubctr}{-1}}}
\def\@@publist{\list
        {[\arabic{pubctr}]\hfill}{\settowidth\labelwidth{[999]}
        \leftmargin\labelwidth
        \advance\leftmargin\labelsep
        \@nmbrlisttrue\def\@listctr{pubctr}}}
 \relax
\makeatother

\def\ba{\begin{equation}}
\def\ea{\end{equation}}

\def\r{\rho}

\def\p{\partial}

\def\no{\noindent}

\def\IR{\relax{\rm I\kern-.18em R}}

\begin{document}


\csname @addtoreset\endcsname{equation}{section}

\newcommand{\eqn}[1]{(\ref{#1})}
\newcommand{\be}{\begin{eqnarray}}
\newcommand{\ee}{\end{eqnarray}}
\newcommand{\non}{\nonumber}
\begin{center}

\vskip -1 cm

\hfill  \\


{\Large \bf Relativistic Superluminal Neutrinos  }

\vskip .7 in

{\bf Alex Kehagias}\phantom{x} 

\vskip 0.2in

Physics Division, National Technical University of Athens, \\
15780 Athens,  Greece\\
{\footnotesize{\tt kehagias@central.ntua.gr}}



\end{center}

\vfill
\vskip .4in

\centerline{\bf Abstract}
We present a possible solution to the OPERA collaboration anomaly for the speed of neutrinos, based on the idea that 
it is a local effect caused by a scalar field sourced by the earth.  The coupling of the 
scalar to neutrinos effectively changes the background metric where 
they propagate, leading to superluminality. The strength of the coupling  is set by a new mass scale, which is at
$1\, {\rm TeV}$ to account for the OPERA anomaly. Moreover, if this scenario is valid, the neutrino velocity depends on the baseline 
distance between the emission and detection points in such a way that superluminal signals are turn to subluminal for baseline 
distances roughly larger than the earth radius.

\no

\vfill
\no

\vskip .5cm

\vfill



\baselineskip 15 pt
\no

The OPERA collaboration \cite{opera} has recently announced evidence for superluminal propagation of $\mu$ neutrino $\nu_\mu$. 
In particular, the reported 
value of the $\mu$ neutrino velocity is
\be
u=1+(2.48\pm 0.28({\rm stat})\pm 0.30 ({\rm sys}))\times 10^{-5}
\ee
i.e. neutrinos are faster that light ($c=1$)! As provocative as it sounds, this measurement 
kindle fantasies of Lorenz symmetry violations and attracted a lot of attention.   

It seems that there are  three possibilities for explaining this result: I) The anomaly is due to some systematic error and the neutrino fly 
at the speed of light. II) The anomaly is real and indicates violation of Lorentz invariance. III) The anomaly is real, neutrinos
fly faster than light but still Lorentz invariance is not violated. 

A reasonable possibility would be to abandon Lorentz symmetry and adopt an energy-depended speed of the $\nu_\mu$ \cite{Amelino}. 
This could explain the reported anomaly. However, in our opinion, it is worth looking for Lorentz symmetric sources of
superluminal neutrino velocity before exploring Lorentz violating ones.   
 
We  consider the OPERA  result as pointing towards the third possibility,  first  because it is a 
carefully designed experiment, and second Lorentz symmetry is a deep rooted postulate for which we are biased.
Adopting possibility III, means that 
we should look for possible neutrino interactions which produce the anomaly. The obvious source is the earth itself 
and its gravitational 
field. However, gravitational effects are tiny and give contributions which for the case of earth is of order
 $R_{S_\oplus}/R_\oplus=10^{-10}$ where $R_\oplus=6.4 \times 10^8 \,{\rm cm}$, where $R_{S_\oplus}=0.886\, {\rm cm}$ 
are the Earth radius and its Schwarzschild radius respectively. This is a tiny correction which cannot 
account for the neutrino speed anomaly and more importantly, it is in the wrong direction as produces delay
in arrival time. To produce advanced arrival of neutrino signals, a kind of anti-gravity or repulsion is needed. 
This is for example the effect of a cosmological constant. In fact, a positive cosmological constant of the order of 
$\Lambda=10^{-21}\, {\rm cm}$ may produce the reported OPERA result. The only problem is that such a cosmological constant 
is too  large (although only 35 orders of magnitude off).  
 Another possibility is that the anomaly is due to  a second spin-2  particle, massive this time,
 coupled to neutrinos \cite{Dvali}. Here we will employ a second possibility, 
namely coupling of neutrinos to a scalar field.

In this direction, let us consider a scalar field $\pi$ with derivative couplings to the neutrino kinetic term the dynamics of 
which is described by the following effective Lagrangian
\be
{\cal{L}}_\nu=\bar{\nu}\gamma^\mu\partial_\mu\nu +L_*^4 \p_\mu \pi\p^\rho\pi\, \bar{\nu}\gamma^\mu\p_\rho\nu  
\label{lang}
\ee 
Here $L_*$ (or $M_*=1/L_*$) is a new scale which specifies the strength of the $\pi$ coupling (the same way Planck length 
specifies the gravitational
strength). One may recognize in $\pi$ a galileon type of scalar field \cite{g1}-\cite{g2}. In the absence of the $\pi$ field, the neutrino satisfies 
\be
\gamma^\mu\partial_\mu \nu=0
\ee
from where the wave-equation
\be
\partial_\mu\partial^\mu \nu=0
\ee  follows. Thus, neutrinos travel at the speed of light $c$ in the absence of scalar couplings. 
Let us now consider the case of a non-zero $\pi$-coupling. In this case, the spherical symmetric static
$\pi$ background field  would be
\be
\pi=\frac{\alpha}{r}
\ee
where $\alpha$ is a numerical parameter\cite{Nicolis:2004qq},\cite{DH},\cite{DK}. We may assume that  $\pi$ is
 sourced by the trace of energy-momentum tensor ${T^\mu}_\mu$ 
so that in the static case 
\be
\alpha=\frac{M}{M_*}\, .
\ee
The sign of $\alpha$ is irrelevant here as it appears quadratically below.
The Dirac equation for the neutrino is in this case
\be
\gamma^\mu\p_\mu\nu+L_*^4\p^\mu\pi\p^\rho\pi\gamma_\mu\,\p_\rho\nu=0
\ee
For the spherical symmetric $\pi$ field we get that 
\be
\gamma^0\partial_t\nu+\gamma^1(\p_\r +\frac{2}{r})\nu+r\slashed D_2\nu+L_*^4\pi'^2\gamma^1(\p_r +\frac{2}{r})\nu=0
\ee
where $\slashed D_2$ is the Dirac operator on $S^2$. From the above expression follows that neutrinos see an effective veilbein
\be
E^t_0=1\, ,~~~E^r_1=1+L_*^4\pi'^2\, ~~~~ E^i_a=r^2 e^i_a\, , ~~~a=3,4\, , ~~i=\theta,\phi
\ee
where $e^i_a$ are the $S^2$ zweinbein. This means that the effective spacetime on which neutrinos propagate  has the metric  
\be
ds^2=-dt^2+\frac{dr^2}{1+L_*^4\,\pi'^2}+r^2 d\Omega_2^2 \label{met}
\ee
It is straightforward then to find that neutrinos travel with velocity $u>c$ 
\be
u^2=1+\pi'^2=1+L_*^4\frac{\alpha^2}{r^4}
\ee
Thus attributing the neutrino speed anomaly to such a coupling, we find that 
\be
(u-c)/c=\beta=\frac{1}{2}L_*^4\pi'^2= 10^{-5}
\ee
so that we get 
\be
2\times 10^{-5}= L_*^6 \frac{M^2}{r^4}
\ee
Then for OPERA we find 
\be
L_*^6=8\times 10^{-5}\frac{L_P^4 R_\oplus^4}{2 R_{S_\oplus}^2}
\ee
which gives a length scale 
\be
L_*=2\times 10^{-17}{\rm cm}
\ee
Surprising, this length corresponds to a mass $M_*$ at the electroweak scale
\be
M_*=1 \, {\rm TeV}
\ee 

By a simple geometric setup one may easily calculate the flying time of neutrinos between two points $A$ and $ B$ on earth at 
baseline distance $d$.
The result is
\be
t_{AB}=\frac{R_\oplus^2}{d}\left(\sqrt{1+2\beta}-\sqrt{2\beta+\left(1-\frac{d^2}{R_\oplus^2}\right)}\right)
\ee 
For small $\beta=10^{-5}$ we get 
\be
t_{AB}=d -\frac{d}{1-\frac{d^2}{R^2}}\beta \label{tab}
\ee
For OPERA, $d=730 \, {\rm km}$ and $d/R\approx 0.1$ so that we have approximately 
\be
t_{AB}=d(1-\beta) 
\ee
as expected.
It should be noted that the neutrino flying time or its velocity depends on the baseline 
distance between the points $A,B$ and the  anomaly in flying time is
\be
\delta t_{AB}=10^{-5} \frac{d}{1-\frac{d^2}{R_\oplus^2}}
\ee
Note that for $d>R_\oplus$, the $\delta t_{AB}$ becomes negative and neutrino arrive delayed. Thus for surface distance 
$s>\pi R_\oplus/3 \approx 6692{\rm km}$, the superluminal speed of neutrinos turns to subluminal.
 Thus if the CERN CNGS beam had to be 
detected at Super-Kamiokande more that $9,000 {\rm km}$ away, a subluminal speed of neutrinos would have been reported.  

One may worry about the supernovae constraints since  observations  of light and neutrinos from the supernovae  
provided strict limits on the
speed of electron (anti) neutrinos. In the
$10 \,{\rm MeV}$ range, a limit of $|u-c|/c < 2 \times 10^{-9}$ has been set by the observation of anti-neutrinos burst 
from the SN1987A supernova \cite{sup}. In fact the equivalence
principle requires that photons and neutrinos to follow the
same geodesic path from SN1987A to Earth. If neutrinos effectively move in the metric (\ref{met}), the same should do the photons 
if equivalence principle is to be respected. This is possible if photons couple to the scalar field much the same way as neutrinos.
For example a coupling of the form
\be
L_{ph}=-\frac{1}{4} F_{\mu\nu}F^{\mu\nu}-\frac{L_*^4}{2} \p^\mu\pi\p^\nu \pi\, F_{\mu\rho}{F_\nu}^\rho
\ee
would produce consistency with supernova result.

Concluding, we have shown here how a non-minimal effective coupling of a scalar to neutrinos may explain the announced advanced arrival of
neutrinos from the  CERN CNGS beam to the Gran Sasso Laboratory  over a baseline of about $730 \, {\rm km}$. The proposed effective 
theory respects Lorentz  symmetry and superluminality is due to interactions, which effectively modifies the spacetime metric 
on which neutrinos propagate  without need of violating fundamental laws. 
In particular, the neutrino speed depends on the baseline distance  according to eq.(\ref{tab}) 
 and it might be possible to search for superluminal to subluminal transition. 

A final comment concerns the provocative value of $M_*$ at the ${\rm TeV}$ scale. This points towards the possibility $\pi$ 
to be a  Standard Model or beyond the Standard Model field, and the Lagrangian (\ref{lang}) to describe a non-renormalizable interaction of this mode
to neutrinos.  

\vskip.2in 
\noindent
I would like to thank Manolis Dris for turning my attention to the OPERA collaboration result and Gia Dvali for discussions.
This work is supported
by the PEVE-NTUA-2009 program.

\end{document}